\documentclass[12pt]{article}
\usepackage{amsmath}
\usepackage{amsfonts} 
\usepackage{graphicx,psfrag,epsf}
\usepackage{enumerate}
\usepackage{natbib}
\usepackage{url} 
\usepackage{tabularx} 
\usepackage{multirow} 
\usepackage{stackrel} 
\usepackage[font={small}]{caption} 
\usepackage{setspace}

\newcommand{\blind}{1}

\newcommand\independent{\protect\mathpalette{\protect\independenT}{\perp}}
\def\independenT#1#2{\mathrel{\rlap{$#1#2$}\mkern2mu{#1#2}}}

\allowdisplaybreaks

\begin{document}

\def\spacingset#1{\renewcommand{\baselinestretch}%
{#1}\small\normalsize} \spacingset{1}


\if1\blind
{
  \title{\vspace*{-35pt} \bf Bayesian Non-Parametric Factor Analysis for Longitudinal Spatial Surfaces} 
  \date{}
  \author{Samuel I.\ Berchuck\thanks{Samuel I.\ Berchuck is a Postdoctoral Associate, Department of Statistical Science and Forge, Duke University, NC 27708 (E-mail: sib2@duke.edu). Mark Janko is a Research Scientist, Institute of Health Metrics and Evaluation, University of Washington, Seattle, WA 98121 (E-mail: mjanko@uw.edu). Felipe A. Medeiros is a Professor, Department of Ophthalmology, Duke University, Durham, NC, 27708 (E-mail: felipe.medeiros@duke.edu). William Pan is an Associate Professor, Duke Global Health Institute, Duke University, Durham, NC 27710 (E-mail: william.pan@duke.edu). Sayan Mukherjee is a Professor, Departments of Statistical Science, Mathematics, Computer Science, and Bioinformatics \& Biostatistics, Duke University, NC, 27708 (E-mail: sayan@stat.duke.edu). This work was partially supported by the National Aeronautics and Space Administration (MJ and WP; NNX15AP74G S005), the National Institutes of Health/National Eye Institute (FAM; EY029885, EY027651, and EY021818), the Human Frontier Science Program (SM; RGP005), and the National Science Foundation (SM; DMS 17-13012, DBI 1661386, and DEB 1840223) as well as high-performance computing partially supported by grant 2016-IDG-1013 from the North Carolina Biotechnology Center (SM).}, Mark Janko, Felipe A. Medeiros, \\William Pan, and Sayan Mukherjee}
  \maketitle
} \fi

\if0\blind
{
  \bigskip
  \bigskip
  \bigskip
  \begin{center}
    {\LARGE \vspace*{75pt} \bf Bayesian Non-Parametric Factor Analysis for Longitudinal Spatial Surfaces}
\end{center}
  \medskip
} \fi

\vspace*{-20pt}

\begin{abstract}
We introduce a Bayesian non-parametric spatial factor analysis model with spatial dependency induced through a prior on factor loadings. For each column of the loadings matrix, spatial dependency is encoded using a probit stick-breaking process (PSBP) and a multiplicative gamma process shrinkage prior is used across columns to adaptively determine the number of latent factors. By encoding spatial information into the loadings matrix, meaningful factors are learned that respect the observed neighborhood dependencies, making them useful for assessing rates over space. Furthermore, the spatial PSBP prior can be used for clustering temporal trends, allowing users to identify regions within the spatial domain with similar temporal trajectories, an important task in many applied settings. In the manuscript, we illustrate the model's performance in simulated data, but also in two real-world examples: longitudinal monitoring of glaucoma and malaria surveillance across the Peruvian Amazon. The R package \texttt{spBFA}, available on \texttt{CRAN}, implements the method.
\end{abstract}

\noindent%

\newpage
\spacingset{1.45} 


\section{INTRODUCTION}
\label{sec:intro}

The covariance for the standard Bayesian factor model, $\boldsymbol{\Psi} = \boldsymbol{\Lambda}\boldsymbol{\Lambda}^\top + \boldsymbol{\Sigma}$, is a matrix decomposition, constructed to learn a latent representation for some potentially high-dimensional data object $\mathbf{Y}_t = \{Y_t(\mathbf{s}_1), \ldots, Y_t(\mathbf{s}_m)\}^\top$. We use notation from the spatial statistics literature to indicate the dimension of $\mathbf{Y}_t$, however this is only for consistency throughout the remainder of the paper. In fact, the data object $\mathbf{Y}_t$ is often not spatial in nature, but a vector that contains a large number of highly collinear variables. As such, throughout this paper, we refer to this dimension as the ``variable dimension" of the data. The subscript $t$ describes observed repetitions of the data object and can be inherently independent, spatial, or temporal in nature; we refer to this data dimension as the ``replication dimension". 

In this manuscript, we deal with the data setting where the vector $\mathbf{Y}_t$ represents a spatial surface and is observed longitudinally across time, $t$. Our ultimate goal is to obtain a low-dimensional representation of $\mathbf{Y}_t$, at each time $t$, that is learned from a process that accounts for the spatial structure of the observed data. By incorporating these spatial dependencies, the hope is that meaningful latent factors are learned that aid in understanding rates of change across the spatial surface and provide a framework for clustering spatial locations based on comparable temporal trajectories. To accomplish this, we generalize the standard factor analysis, to allow for non-linear relationships (Equation \ref{eq:lik}) and introduce a novel spatial Bayesian non-parametric (BNP) prior on the columns of the factor loadings matrix, $\boldsymbol{\Lambda}$ (Equation \ref{eq:psbp}). We begin by reviewing existing factor analysis methods for spatial data.

Factor analysis is characterized by dimension reduction along the variable dimension of the observed data and is accomplished by projecting the data into a lower dimensional space, defined by a set of $k$ factors, $\boldsymbol{\eta}_t = (\eta_{t1},\ldots, \eta_{tk})^\top$. In practice the number of factors is small compared to the dimension of the data object ($k \ll m$). By definition, the factors have lower variability than the original data and are more manageable for inferential purposes due to their low dimension. In a standard Bayesian factor analysis, the latent vector $\boldsymbol{\eta}_t$ is often modeled as a standard Gaussian \citep{murray2013bayesian}. 

Typically, much of the innovation in factor analysis involves the prior for the $m \times k$ dimensional factor loadings matrix, $\boldsymbol{\Lambda}$. The naive approach assumes independent Gaussian priors for each element of $\boldsymbol{\Lambda}$, which has obvious computational issues when $m$ and $k$ are large. Furthermore, it may lead to poor inference due to the weakness of the prior specification and is not identifiable without further restrictions. In general, the specification of $\boldsymbol{\Psi}$ is not unique, as there are infinitely many possible factor loading matrices that satisfy the form. This can be seen by noting that any matrix of the form $\boldsymbol{\Lambda} \mathbf{P}$ satisfies the condition, for any orthogonal matrix $\mathbf{P}$ (i.e., $\mathbf{P}\mathbf{P}^\top = \mathbf{I}_k$). 

To remedy this, $\boldsymbol{\Lambda}$ is often a lower diagonal matrix with the loadings on the diagonal forced to be positive. This has been made computationally more efficient in recent years through parameter expansion of the loadings using basis elements \citep{ghosh2009default}. Although these methods can be useful for identifiability, they remain computationally burdensome. Furthermore, it has been noted that from a Bayesian perspective, one does not require identifiability for many applications, including prediction, covariance estimation, and clustering \citep{bhattacharya2011sparse}. 

Adaptations of factor analysis to the spatial setting are plentiful and predominately focus on spatial dependence in the replication dimension of the data. A typical application of spatial factor analysis involves learning a latent representation of some high-dimensional data object that is observed across a geography, whether point-referenced or areal. Here, the foundational assumption of spatial statistics, that dependence between observations weakens as the distance between locations increases, is applied to the factors, so that a latent factor at a location should be similar to factors at nearby locations. 

\cite{christensen2002latent} used this assumption to fit a shift-factor analysis method to model multivariate spatial data with temporal behavior modeled by autoregressive (AR) components. This method entertained several forms of spatial dependence through a single factor, a standard construction in the literature \citep{hogan2004bayesian}. There have been many extensions to multiple factors, most often using Gaussian likelihoods \citep{nethery2015common}, but also generalizing to Poisson \citep{tzala2008bayesian} and binary \citep{wall2009spatial}. There are also extensions to informative missingness \citep{reich2010latent}, and spatial mis-alignment \citep{nethery2018joint}. 

In all of these methods, the latent factors are responsible for encoding spatial dependency for the purpose of reducing the observed data at each location. In this paper, however, we will focus on an alternative form of spatial factor analysis that instead introduces spatial structure along the variable dimension of the data. Thus, instead of dimension reduction for some high-dimensional response across locations, the response is now univariate, and the dimension reduction is performed across spatial units. This approach is advantageous when the modeling goal is to identify spatial clusters whose temporal behavior is similar. 

This approach was introduced by \cite{lopes2008spatial} through a spatial dynamic factor model. The key to this approach is a spatial prior on the columns of the factor loadings matrix, that allows for dimension reduction to be informed by spatial proximities. Space was modeled using a distance-based Gaussian random field, while a more recent version uses a Gaussian Markov random field for sparsity purposes \citep{strickland2011fast}. This method has been extended to the generalized likelihood setting \citep{lopes2011generalized}. These methods use a lower diagonal specification for the loadings matrix for identifiability purposes and the number of factors is learned through reversible jump Markov chain Monte Carlo (MCMC). While these methods are useful for learning factors across a spatial surface, they rely on complicated identifiability constraints and lack clustering properties. 

In this manuscript, we introduce a spatial factor analysis that collapses spatial locations into meaningful latent factors using a spatial BNP prior for the fully specified factor loadings matrix. The method yields a non-separable and non-stationary spatiotemporal (ST) process, with temporal dependence introduced through the factors. We show that the BNP prior offers benefits for scalability and is useful for clustering spatial locations into regions across space with similar risk trajectories. A computationally efficient MCMC sampler is introduced that uses slice sampling to allow for an infinite mixture model. Furthermore, a multiplicative gamma process shrinkage prior is used to adaptively determine the number of latent factors, avoiding the computational intensive reversible jump technique.

This paper is outlined as follows. In Section \ref{sec:methods}, we introduce a general factor analysis modeling framework and detail our novel spatial BNP prior for the columns of the factor loadings matrix. Through simulation in Section \ref{sec:sims}, we assess the utility of the novel prior in ST data and for clustering temporal trends across a spatial surface. Then, in Section \ref{sec:data}, we apply the model to two real-world data applications: glaucoma disease progression and malaria risk surveillance. We conclude in Section \ref{sec:sum} with a discussion.


\section{METHODOLGY}
\label{sec:methods}

We begin by introducing a generalized modeling framework for factor analysis that allows for non-linearity and detail the temporal process for the latent factors. We then introduce the spatial BNP prior for the factor loadings matrix and describe the multiplicative gamma process shrinkage prior for adaptively learning the appropriate number of latent factors. We conclude the section by working out an MCMC sampler for the infinite mixture model, describing the clustering properties of the introduced prior, and detailing prediction theory.


\subsection{A General Modelling Framework}
\label{sec:model}

A generalized factor analysis model can be written as follows,
\begin{align} \label{eq:lik}
Y_{t}(\mathbf{s}_{i,o}) | \vartheta_{t}(\mathbf{s}_{i,o}), \boldsymbol{\zeta}_{t}(\mathbf{s}_{i,o}) &\stackrel{\text{ind}}{\sim}{} f\left(Y_{t}(\mathbf{s}_{i,o}); g^{-1}\left(\vartheta_{t}(\mathbf{s}_{i,o})\right), \boldsymbol{\zeta}_{t}(\mathbf{s}_{i,o})\right)\\
g\left(\vartheta_{t}(\mathbf{s}_{i,o})\right) &= \mathbf{x}_{t}(\mathbf{s}_{i,o})\boldsymbol{\beta} + \sum_{j = 1}^k \lambda_{j}(\mathbf{s}_{i,o}) \eta_{tj}. \notag
\end{align}
Here, we formally define our observed data as $Y_{t}(\mathbf{s}_{i,o})$ for temporal visit $t$, $(t = 1,\ldots,T)$ and spatial realization $\mathbf{s}_{i,o}$, for location $i$, $(i = 1,\ldots,m)$, and observation type $o$, $(o = 1,\ldots, O)$. This is a general specification, so that at each time $t$, the spatial object can be multi-layered (i.e., color channels or multiple disease outcomes per location) with $O$ layers. We define vectorized versions of the observed data as follows, $\mathbf{Y} = (\mathbf{Y}_{1}^{\top}, \ldots, \mathbf{Y}_T^{\top})^\top$, where $\mathbf{Y}_t = \left(\mathbf{Y}_{t1}^\top,\ldots,\mathbf{Y}_{tO}^\top\right)^\top$ and $\mathbf{Y}_{to} = \{Y_{t}(\mathbf{s}_{1,o}), \ldots, Y_{t}(\mathbf{s}_{m,o})\}^\top$. 

In our specification the factor loadings matrix is fully specified, with loadings $\lambda_j(\mathbf{s}_{i,o})$, corresponding to the stacking of the observed data. So the $j^{th}$ column is given by $\boldsymbol{\lambda}_j = (\boldsymbol{\lambda}_{j1}^\top,\ldots,\boldsymbol{\lambda}_{jO}^\top)^\top$, with $\boldsymbol{\lambda}_{jo} = \{\lambda_j(\mathbf{s}_{1,o}), \ldots, \lambda_j(\mathbf{s}_{m,o})\}^\top$. A full specification allows for a direct application of spatial structure to $\boldsymbol{\lambda}_j$, $j = 1,\ldots,k$, as it has the same dimension as the underlying process, $mO$. While a full specification limits the interpretability of the factors themself, as mentioned before, one does not require identifiability for many applications.

The scalar form of the likelihood in Equation \ref{eq:lik} motivates spatial dependency in the factor loadings. In particular, due to the fully specified factor loadings matrix, the following linear relationship between the transformed mean process and the latent factors exists,
\begin{equation} \label{eq:mean}
g\left(\vartheta_{t}(\mathbf{s}_{i,o})\right) = \mathbf{x}_{t}(\mathbf{s}_{i,o})\boldsymbol{\beta} + \sum_{j = 1}^k \lambda_{j}(\mathbf{s}_{i,o}) \eta_{tj} = \mathbf{x}_{t}(\mathbf{s}_{i,o})\boldsymbol{\beta} + \lambda_{1}(\mathbf{s}_{i,o}) \eta_{t1} + \cdots + \lambda_{k}(\mathbf{s}_{i,o}) \eta_{tk}
\end{equation}
This illuminates that a factor loading $\lambda_{j}(\mathbf{s}_{i,o})$ represents the amount that observation $Y_{t}(\mathbf{s}_{i,o})$ is explained through the latent factor $j$ at time $t$, $\eta_{tj}$. Therefore, for two observations, $Y_{t}(\mathbf{s}_{i,o})$ and $Y_{t}(\mathbf{s}_{i',o})$, that are spatially correlated, we would assume that their relationships to the latent factor, $\eta_{tj}$ would be similar, $\lambda_{j}(\mathbf{s}_{i,o}) \approx \lambda_{j}(\mathbf{s}_{i',o})$. 

While standard Bayesian factor analysis is performed using a Gaussian likelihood, Equation \ref{eq:lik} is a generalized form. The Gaussian specification can be recovered if we choose $f$ to be Gaussian with mean, $\mu_{t}(\mathbf{s}_{i,o}) = g^{-1}\left(\vartheta_{t}(\mathbf{s}_{i,o})\right)$, nuisance or variance, $\boldsymbol{\zeta}_{t}(\mathbf{s}_{i,o}) = \sigma^2(\mathbf{s}_{i,o})$ and $g$ the identity link. This is equivalent to the following vectorized model specification, $\mathbf{Y}_t = \mathbf{X}_t\boldsymbol{\beta} + \boldsymbol{\Lambda} \boldsymbol{\eta}_t + \boldsymbol{\epsilon}_t$, with $\boldsymbol{\epsilon}_t \sim \text{N}_{mO}(\mathbf{0},  \boldsymbol{\Sigma}),$ where  $\boldsymbol{\Sigma} = \text{Diag}(\boldsymbol{\sigma}_1^2, \ldots, \boldsymbol{\sigma}_O^2)$, and $\boldsymbol{\sigma}_o^2 = (\sigma^2(\mathbf{s}_{1,o}), \ldots, \sigma^2(\mathbf{s}_{m,o}))^\top$. 
 The component, $\mathbf{X}_t\boldsymbol{\beta}$, allows for covariates to adjust the factor analysis. The design matrix, $\mathbf{X}_t$, has rows, $\mathbf{x}_{t}(\mathbf{s}_{i,o})$, which is $p$ dimensional.

The purpose of writing the model in this general form is that it is more flexible, allowing for various likelihoods. For example, when we study malaria in Section \ref{sec:peru} we will model the disease counts as binomial, by specifying $f$ as a binomial distribution, with probability $\pi_{t}(\mathbf{s}_{i,o}) = g^{-1}\left(\vartheta_{t}(\mathbf{s}_{i,o})\right)$, total number of trials fixed, $\boldsymbol{\zeta}_{t}(\mathbf{s}_{i,o}) = n_{t}(\mathbf{s}_{i,o})$ and the logit link.

We conclude this section by specifying a temporal structure for the latent factors. Again, we specify a general framework, $\boldsymbol{\eta} \sim \text{N}\left(\mathbf{0}, \mathbf{H}(\psi) \otimes \boldsymbol{\Upsilon}\right)$, where $\boldsymbol{\eta} = \{\boldsymbol{\eta}_1^\top,\ldots, \boldsymbol{\eta}_{T}^\top\}^\top$. This form is flexible, allowing for many common time series models, including the AR(1) and exponential processes To obtain the AR(1), choose $\mathbf{H}(\psi)$, such that $\left[\mathbf{H}(\psi)\right]_{tt'} = \psi^{|x_t - x_{t'}|}$, which results in $\boldsymbol{\eta}_t = \psi \boldsymbol{\eta}_{t-1} + \boldsymbol{\upsilon}_t, \boldsymbol{\upsilon}_t \sim \text{N}(\boldsymbol{\eta}_{t-1}, \boldsymbol{\Upsilon})$, if time is uniform. The exponential can be obtained with $\left[\mathbf{H}(\psi)\right]_{tt'} = \exp\{\psi |x_t - x_{t'}|\}$, where $x_t$ is follow-up time $t$.


\subsection{Spatial Bayesian Non-parametric Factor Loadings}
\label{sec:psbp}

In order to introduce spatial dependency into the columns of the factor loadings we use 

\noindent the probit stick-breaking process (PSBP), which is a scalable extension of standard spatial processes that allows for clustering. The BNP world has a rich literature involving spatial processes, mainly involving extensions of the Dirichlet Process (DP). The DP is the work-horse of BNPs and, when considering spatial dependencies, is best represented using a stick-breaking construction, such that $G \sim DP(\alpha, G_0)$ if and only if $G(\cdot) = \sum_{l = 1}^{\infty} w_{l}\delta_{\theta_l}(\cdot)$, where $\theta_l \stackrel{\text{iid}}{\sim}{} G_0$ and $w_l = u_l \prod_{r = 1}^{l-1}(1 - u_r), l = 2,3,\ldots$, with $u_r \stackrel{\text{iid}}{\sim}{} \text{Beta}(1, \alpha)$ and $\delta_{\theta_l}$ is a Dirac distribution with point mass at $\theta_l$. Since the introduction of the dependent DP by \cite{maceachern1999dependent}, which modeled dependency through covariate information in the atoms ($\theta_l$) and the weights ($w_l$), many methods have extended the DP to the spatial setting. 

A popular spatial DP extension is \cite{gelfand2005bayesian}, which places a univariate stationary Gaussian process on the atoms to yield a random spatial process that is neither Gaussian nor stationary. The process has been extended to the generalized framework \citep{duan2007generalized}. Modeling spatial dependency through the weights of the stick-breaking representation has also been popular, however until recent years has been computationally inefficient. In the more general stick-breaking construction, \cite{rodriguez2011nonparametric} introduced the PSBP, which replaces the characteristic Beta distribution prior with probit transformations of normal random variables. With the introduction of the PSBP, incorporating spatial dependency in BNP priors has become computationally straightforward, and mainstream \citep{chung2009nonparametric,pati2013posterior,pati2014bayesian}.

To induce the desired spatial dependency, as motivated by Equation \ref{eq:mean}, into the columns of the factor loadings matrix, we use a PSBP for each column,
\begin{align}
\label{eq:psbp}
\lambda_{j}(\mathbf{s}_{i,o}) | G_{j}^{i,o} &\stackrel{\text{ind}}{\sim}{} G_{j}^{i,o}, \quad i=1,\ldots,m, \quad o = 1,\ldots, O, \quad j = 1,\ldots, k \notag\\ 
G_{j}^{i,o}(\cdot) &= \sum_{l=1}^{L} w_{jl}(\mathbf{s}_{i,o}) \delta_{\theta_{jl}}(\cdot)\\
w_{jl}(\mathbf{s}_{i,o}) &= \Phi(\alpha_{jl}(\mathbf{s}_{i,o})) \prod_{r < l} [1 - \Phi(\alpha_{jr}(\mathbf{s}_{i,o}))], \notag
\end{align}
where $\{\alpha_{jl}(\mathbf{s}) : \mathbf{s} \in \mathcal{D}\}_{l=1}^{L-1}$ for $j = 1,\ldots,k$ has Gaussian marginals, with $\mathcal{D}$ some multivariate spatial surface, and $\{\theta_{jl} \}_{l=1}^L$ for $j = 1,\ldots,k$ are independent and identically distributed for each $j$. The form of Equation \ref{eq:psbp} closely mirrors the stick-breaking construction of the DP, however the weights are now constructed using the standard Gaussian cumulative distribution function, $\Phi$. As is shown in \cite{rodriguez2011nonparametric}, this is a proper construction, because for finite $L$, it ensures that the weights sum to one, and that when $L \rightarrow \infty$ they sum to one almost surely. This property clearly transfers to our new prior across the columns of the factor loadings.

It is useful to interpret the induced conditional likelihood, $f\left(Y_{t}(\mathbf{s}_{i,o}) | G_{j}^{i,o}, \boldsymbol{\eta}_t, \boldsymbol{\zeta}_t(\mathbf{s}_{i,o})\right) = \int \cdots \int f\left(Y_{t}(\mathbf{s}_{i,o}); g^{-1}\left(\vartheta_{t}(\mathbf{s}_{i,o})\right), \boldsymbol{\zeta}_{t}(\mathbf{s}_{i,o})\right) G_{1}^{i,o}(d\lambda_{1}(\mathbf{s}_{i,o})) \cdots G_{k}^{i,o}(d\lambda_{k}(\mathbf{s}_{i,o}))$, which can alternatively be written as $\sum_{l_1 = 1}^L \cdots \sum_{l_k = 1}^L w_{1l_1}(\mathbf{s}_{i,o}) \cdots w_{kl_k}(\mathbf{s}_{i,o}) f\left(Y_{t}(\mathbf{s}_{i,o}); g^{-1}\left(\vartheta_{t}(\mathbf{s}_{i,o})\right), \boldsymbol{\zeta}_{t}(\mathbf{s}_{i,o})\right)$. These two equivalent forms of the induced model demonstrate the mixing, which averages over the factor loadings according to the PSBP (Equation \ref{eq:psbp}). These representations will be useful for determining the marginal and conditional moments in Section \ref{sec:props}.

The parameters that dictate the weights, $\alpha_{jl}(\mathbf{s}_{i,o})$, have a joint distribution that induces spatial dependency. Define the joint parameter, $\boldsymbol{\alpha}_{jlo} = \left\{\alpha_{jl}(\mathbf{s}_{1,o}),\ldots,\alpha_{jl}(\mathbf{s}_{m,o})\right\}^\top$ and $\boldsymbol{\alpha}_{jl} = \{\boldsymbol{\alpha}_{jl1}^\top,\ldots,\boldsymbol{\alpha}_{jlO}^\top\}^\top$. In order to maintain computational feasibility any spatial structure can be specified that can be expressed in a Gaussian kernel, which allows the majority of spatial techniques \citep{banerjee2003hierarchical}. We specify a simple, but flexible form using a non-separable specification, $\boldsymbol{\alpha}_{jl} \sim \text{N}_{Om}\left(\mathbf{0},\boldsymbol{\kappa} \otimes \mathbf{F}(\rho)\right)$. Notice, that while we treat space using a separable process, the resulting marginal process will be non-separable.

The $m \times m$ matrix $\mathbf{F}(\rho)$ dictates the spatial neighborhood structure, for example a Gaussian process with exponential correlation, $\mathbf{F}(\rho) = \exp\{-\rho\mathbf{D}\}$, for a continuous spatial domain or a Gaussian Markov random field for discrete spatial data, $\mathbf{F}(\rho)^{-1} = \mathbf{D}_w - \rho\mathbf{W}$; we assume a proper conditional autoregressive (CAR) prior. Here $\mathbf{D}$ is a distance matrix (typically Euclidean) and $\mathbf{W}$ is an adjacency matrix, with adjacencies $\{w_{ii'}\}$ that indicate the level of spatial correlation between locations $i$ and $i'$ and do not change over time ($\mathbf{D}_w$ is a diagonal matrix that weights the number of neighbors of each locations $i$). The parameter $\rho$ indicates the level of spatial correlation.

Finally, in prior attempts to model the factor loadings matrix the number of latent factors (i.e., number of columns of $\boldsymbol{\Lambda}$) was determined using the reversible jump MCMC. This decision requires a preliminary run for each choice of the number of factors and is very computationally intensive. As such, we decide to model the atoms, $\theta_{jl}$, using a multiplicative gamma process shrinkage prior \citep{bhattacharya2011sparse}. This prior conveniently shrinks the magnitude of possible entries, where the degree of shrinkage increases with the column index. In particular, $\theta_{jl} \stackrel{\text{ind}}{\sim}{} \text{N}(0, \tau_j^{-1})$, where the precision is forced to increase over the column index, $\tau_j = \prod_{h=1}^j \delta_h,$ with $\delta_1 \sim \text{Ga}(a_1, 1)$, and $\delta_h \sim \text{Ga}(a_2, 1)$, for $h \geq 2$. This allows us to specify a value of $k$ that is larger than the number of supposed factors, with the prior reducing the factors to a set of meaningful ones.


\subsection{Computational Considerations}
\label{sec:comp}

In order to facilitate Bayesian inference, the likelihood can be written in terms of the underlying atoms, a standard practice in mixture models,
\begin{align}
\label{eq:disc}
g\left(\vartheta_{t}(\mathbf{s}_{i,o})\right) = x_{t}(\mathbf{s}_{i,o})\boldsymbol{\beta} + \sum_{j = 1}^k \theta_{j \xi_{j}(\mathbf{s}_{i,o})} \eta_{tj}.
\end{align}
This is a simple replacement of the factor loadings, $\lambda_{j}(\mathbf{s}_{i,o})$, with their corresponding atom, $\theta_{j l}$, which is determined by a clustering indicator $\xi_{j}(\mathbf{s}_{i,o}) = l$. The representation in Equation \ref{eq:disc} reminds us of the discrete nature of the PSBP as the categorical parameter, $\xi_{j}(\mathbf{s}_{i,o})$, indicates the cluster of $\lambda_{j}(\mathbf{s}_{i,o})$, and has the following distribution, $\text{Multinomial}(w_{j1}(\mathbf{s}_{i,o}), \ldots, w_{jL}(\mathbf{s}_{i,o}))$, so that $P(\xi_{j}(\mathbf{s}_{i,o}) = l) = w_{jl}(\mathbf{s}_{i,o})$. This construction helps to illuminate the importance of the spatial dependency introduced in the PSBP, as the value of $\xi_{j}(\mathbf{s}_{i,o})$ (i.e., cluster of $\lambda_{j}(\mathbf{s}_{i,o})$) is sampled from a multinomial distribution with weights that have been spatially smoothed to be similar to nearby locations. This is desirable, because it encourages close locations to belong to the same cluster, and thus constructs underlying factors that relate to regions of the spatial domain.

We can further facilitate efficient computations by introducing the latent variable, $z_{jl}(\mathbf{s}_{i,o}) \stackrel{\text{ind}}{\sim}{} \text{N}(\alpha_{jl}(\mathbf{s}_{i,o}), 1)$. If we take $\xi_{j}(\mathbf{s}_{i,o}) = l$ if and only if $z_{jl}(\mathbf{s}_{i,o}) > 0$ and $z_{jr}(\mathbf{s}_{i,o})$ for $r < l$, we get the following property, $P(\xi_{j}(\mathbf{s}_{i,o}) = l | z_{jl}(\mathbf{s}_{i,o})) = P(z_{jl}(\mathbf{s}_{i,o}) > 0, z_{jr}(\mathbf{s}_{i,o}) < 0,\forall r < l) = \Phi(\alpha_{jl}(\mathbf{s}_{i,o})) \prod_{r < l} [1 - \Phi(\alpha_{jr}(\mathbf{s}_{i,o}))] = w_{jl}(\mathbf{s}_{i,o}).$ This permits conjugacy in the $\alpha_{jl}(\mathbf{s}_{i,o})$ by noting the following conditional independence, $\xi_{j}(\mathbf{s}_{i,o}) \independent \alpha_{jl}(\mathbf{s}_{i,o}) | z_{jl}(\mathbf{s}_{i,o})$. Furthermore, the data augmentation parameters, $z_{jl}(\mathbf{s}_{i,o})$, have conjugate form.

The theory described above was for finite $L$, however it can be easily extended to an infinite mixture model using a slice sampling technique \citep{walker2007sampling}. Slice sampling makes the infinite mixture model computationally feasible by introducing an upper bound for the number of clusters, thus reducing the process to a finite mixture model. In particular, all parameters that depend on the number of clusters ($\theta_{jl}$, $\xi_{j}(\mathbf{s}_{i,o})$, $z_{jl}(\mathbf{s}_{i,o})$, $\alpha_{jl}(\mathbf{s}_{i,o})$, $\boldsymbol{\kappa}$, $\rho$, $\delta_h$) will be augmented using the slice sampling truncation. The idea is to introduce a latent variable, $u_{j}(\mathbf{s}_{i,o})$, with uniform density so that conditional on $u_{j}(\mathbf{s}_{i,o}), j = 1,\ldots, k$, $o = 1,\ldots, O$ and $i = 1,\ldots, m$, the conditional mixture distribution becomes finite. When dealing with full conditionals, this truncation corresponds to reducing the number of mixture components for each column of the factor loadings matrix to $L_j^*$, so that $l_j = 1,\ldots, L_j^* = \max\{L_{j}^{i,o}; i = 1,\ldots,m, o = 1,\ldots,O\}$, where $L_{j}^{i,o}$ is the minimum integer satisfying $\sum_{l_j = 1}^{L_{j}^{i,o}} w_{jl_j}(\mathbf{s}_{i,o}) > 1 - u_j^* = \min\{u_{j}(\mathbf{s}_{i,o})\}$, for $i = 1,\ldots, m$ and $o = 1,\ldots,O$. Throughout the simulations and data illustrations in Sections \ref{sec:sims} and \ref{sec:data}, we use the infinite mixture model. Full computational details are in the Supplementary Materials.


\subsection{Model Properties} 
\label{sec:props}

We focus on the class of spatial PSBP models, where each column of the factor loadings matrix has the following form, $\mathcal{M}_j = \{G_{j}^{i,o} : \mathbf{s}_{i,o} \in \mathcal{D}\}$, where each column progressively shrinks due to the gamma process shrinkage prior on the atoms. For each column, the process $G_{j}^{i,o}$ marginally follows a PSBP for each $\mathbf{s} \in \mathcal{D}$. Therefore, for any set $B \in \mathcal{B}$, we can obtain the moments of the process. In this section, we describe the moments of the PSBP process, originally derived in \cite{rodriguez2011nonparametric}, plus new results that describe the conditional and marginal moments of the introduced spatial factor analysis. 

The process moments are as follows, beginning with the mean, $\mathbb{E}[G_{j}^{i,o}(B)] = G_{0j}(B)$. The variance, $\mathbb{V}(G_{j}^{i,o}(B))$, and covariance, $\mathbb{C}(G_{j}^{i,o}(B), G_{j}^{i',o'}(B))$, are as follows,
\begin{align} \label{eq:psbp_moments}
\begin{split}
&G_{0j}(B)\{1 - G_{0j}(B)\} \left[\beta_{2}(\mathbf{s}_{i,o})\left( \frac{1 - \{1 - 2 \beta_{1}(\mathbf{s}_{i,o}) + \beta_{2}(\mathbf{s}_{i,o})\}^L}{2\beta_{1}(\mathbf{s}_{i,o}) - \beta_{2}(\mathbf{s}_{i,o})} \right) \right]\\
&G_{0j}(B)\{1 - G_{0j}(B)\} \beta_{2}(\mathbf{s}_{i,o}, \mathbf{s}_{i',o'}) \left[\frac{1 - \{1 - \beta_{1}(\mathbf{s}_{i,o}) - \beta_{1}(\mathbf{s}_{i',o'}) + \beta_{2}(\mathbf{s}_{i,o},\mathbf{s}_{i',o'})\}^L}{\beta_{1}(\mathbf{s}_{i,o}) + \beta_{1}(\mathbf{s}_{i',o'}) - \beta_{2}(\mathbf{s}_{i,o},\mathbf{s}_{i',o'})}\right]
\end{split}
\end{align}
Finally, it is easy to see that the covariance across columns is zero, meaning the shrinkage process does not introduce dependency at the PSBP level, $\mathbb{C}[G_{j}^{i,o}(B), G_{j'}^{i,o}(B)] = G_{0j}(B)G_{0j'}(B)\{1 - (1 - \beta_{1}(\mathbf{s}_{i,o}))^L\}^2 - G_{0j}(B)G_{0j'}(B) \stackrel{L \rightarrow \infty}{=}{} 0.$ Here, we use the specification that, $u_{jl}(\mathbf{s}_{i,o}) = \Phi(\alpha_{jl}(\mathbf{s}_{i,o}))$, $\beta_{p}(\mathbf{s}_{i,o}) = \mathbb{E}[u_{jl}(\mathbf{s}_{i,o})^p]$ and the base distribution for atom, $\theta_{jl}$, is given by $G_{{0j}}$.  Higher moments are also defined as such, $\beta_{2}(\mathbf{s}_{i,o}, \mathbf{s}_{i',o'}) = \mathbb{E}[u_{jl}(\mathbf{s}_{i,o}) u_{jl}(\mathbf{s}_{i',o'})].$ 

As described in \cite{rodriguez2011nonparametric}, these stick-breaking expectations, which show up in the model properties, have closed forms as long as the underlying spatial process has a marginal distribution, $\alpha_{jl}(\mathbf{s}_{i,o}) \sim \text{N}(\mu,\sigma^2).$ Then, using a change of variables, $t_1 = \alpha_{jl}(\mathbf{s}_{i,o}) - x$ and $t_2 = \alpha_{jl}(\mathbf{s}_{i,o})$, we see that $\beta_{1}(\mathbf{s}_{i,o}) = P(T_1 > 0)$, where $T_1 \sim \text{N}(\mu, 1 + \sigma^2)$. Generally, the  $p$-th moment is given by, $\beta_{p}\left(\mathbf{s}_{i,o}\right) = \mathbb{E}\left[u_{jl}^p(\mathbf{s}_{i,o})\right] = P(T_1 > 0, \ldots, T_p > 0)$, where $(T_1,\ldots,T_p)^\top$ is multivariate normal with $\mathbb{E}[T_i] = \mu$, $\mathbb{V}(T_i) = 1 + \sigma^2$, and $\mathbb{C}(T_i,T_j) = \sigma^2$. Finally, $\beta_{2}(\mathbf{s}_{i,o}, \mathbf{s}_{i',o'}) = P(T_1 > 0, T_2 > 0)$, where $(T_1, T_2)^\top \sim \text{N}(\boldsymbol{\mu},\boldsymbol{\Sigma} + \mathbf{I})$, where the moments $\boldsymbol{\mu}$ and $\boldsymbol{\Sigma}$ come from the marginal distribution, $\left(\alpha_{jl}(\mathbf{s}_{i,o}),\alpha_{jl}(\mathbf{s}_{i',o'})\right)^\top\sim \text{N}(\boldsymbol{\mu},\boldsymbol{\Sigma})$.

The marginal moments indicate the importance of the base distribution, $G_{0j}$, as a centering, with the marginal moments of the underlying Gaussian parameter, $\alpha_{jl}(\mathbf{s}_{i,o})$ controlling the variance and covariance of the sampled distributions around $G_{0j}$. Furthermore, it was shown that as $\mathbf{s}_{i,o} \rightarrow \mathbf{s}_{i',o}$, that $\mathbb{C}(G_{j}^{i,o}(B), G_{j}^{i',o}(B)) \rightarrow \mathbb{V}(G_{j}^{i,o}(B))$, which can be explained by the fact that $\beta_{2}(\mathbf{s}_{i,o}, \mathbf{s}_{i',o}) \rightarrow \beta_1(\mathbf{s}_{i,o}) \beta_1(\mathbf{s}_{i',o})$.

We now turn our attention to the moments of the introduced spatial factor analysis. The following properties are for Gaussian data, as the derivations become untenable without the identity link. The conditional mean and variance are, $\mathbb{E}\left[Y_{t}(\mathbf{s}_{i,o}) | G_{i}^{i,o}, \boldsymbol{\eta}_t, \sigma^2(\mathbf{s}_{i,o})\right] = \mathbf{x}_t(\mathbf{s}_{i,o}) \boldsymbol{\beta} + \sum_{j=1}^k \left(\sum_{l_j = 1}^L  w_{jl_j}(\mathbf{s}_{i,o}) \theta_{jl_j} \right) \eta_{tj}$, and $\mathbb{V}\left(Y_{t}(\mathbf{s}_{i,o}) | G_{j}^{i,o}, \boldsymbol{\eta}_t, \sigma^2(\mathbf{s}_{i,o})\right) = \sigma^2(\mathbf{s}_{i,o}).$ The mean process is elegant, as it takes the form of the original mean process, but replaces the loadings with a mixture over the underlying atoms, weighted according to the PSBP. The spatial covariance, $\mathbb{C}\left(Y_{t}(\mathbf{s}_{i,o}), Y_{t}(\mathbf{s}_{i',o'}) | G_{j}^{i,o}, G_{j}^{i',o'},\boldsymbol{\eta}_t, \sigma^2(\mathbf{s}_{i,o}), \sigma^2(\mathbf{s}_{i',o'})\right)$, is conditionally zero. The form of the conditional moments are reminiscent of a standard spatial analysis that uses a spatially varying intercept with a Gaussian process, where conditional on the spatial intercepts, the process is is independent across space. 

Finally, we turn our attention to the marginal moments of the process. The mean is as follows, $\mathbb{E}\left[Y_{t}(\mathbf{s}_{i,o})\right] = 0$, while the variance, $\mathbb{V}\left(Y_{t}(\mathbf{s}_{i,o})\right)$, and covariance, $\mathbb{C}\left(Y_{t}(\mathbf{s}_{i,o}), Y_{t}(\mathbf{s}_{i',o'})\right)$, are given respectively as,
\begin{align} \notag
&\mathbb{E}\left[\sigma^2(\mathbf{s}_{i,o})\right] + \left[\frac{\beta_{2}(\mathbf{s}_{i,o})\left(1 - \{1 - 2 \beta_{1}(\mathbf{s}_{i,o}) + \beta_{2}(\mathbf{s}_{i,o})\}^L\right)}{2\beta_{1}(\mathbf{s}_{i,o}) - \beta_{2}(\mathbf{s}_{i,o})} \right] \left(\sum_{j=1}^k \tau_j^{-1} \mathbb{E}\left[\eta^2_{tj}\right] \right), \text{ and} \notag\\
&\left[ \frac{\beta_{2}(\mathbf{s}_{i,o}, \mathbf{s}_{i',o'})\left(1 - \{1 - \beta_{1}(\mathbf{s}_{i,o}) - \beta_{1}(\mathbf{s}_{i',o'}) + \beta_{2}(\mathbf{s}_{i,o},\mathbf{s}_{i',o'})\}^L\right)}{\beta_{1}(\mathbf{s}_{i,o}) + \beta_{1}(\mathbf{s}_{i',o'}) - \beta_{2}(\mathbf{s}_{i,o},\mathbf{s}_{i',o'})}\right] \left(\sum_{j=1}^k  \tau_j^{-1} \mathbb{E}\left[\eta^2_{tj}\right]\right).
\end{align}
Both the variance and covariance take the same form as the moments from the PSBP process in Equation \ref{eq:psbp_moments}, however they are now being scaled by a summation, that is a function of the atom variances for each column and the second moment of the latent factors. In particular, as we see increases in both the number of factors ($k$) and the variability in the underlying atoms, the variance and covariance become inflated. For a full interpretation of the marginal moments, we need to place priors on the hyperparameters.


\subsection{Prior Specification}
\label{sec:hypers}

We finalize the model specification by introducing priors for the remaining parameters; spatial parameters, $\boldsymbol{\kappa}$ and $\rho$, temporal parameters, $\boldsymbol{\Upsilon}$ and $\psi$, along with any nuisance parameters, for example the variance in the Gaussian likelihood, $\sigma^2(\mathbf{s}_{i,o})$. For each of these parameters, we choose standard priors to promote conjugacy in the full conditionals. 

The spatial covariance, $\boldsymbol{\kappa}$, is an $O \times O$ matrix over the multiple levels of the image and has the conjugate inverse-Wishart (IW) prior, $\boldsymbol{\kappa} \sim \text{IW}\left(\upsilon, \boldsymbol{\Theta}\right)$. When $O=1$ this prior reduces to an inverse-Gamma distribution, $\text{IG}(\upsilon/2,\boldsymbol{\Theta}/2)$. For degrees of freedom, we specify $\upsilon = O + 1$ and the scale matrix we use $\boldsymbol{\Theta} = \mathbf{I}_O$. This prior is appealing since it induces marginally uniform priors on the correlations of $\boldsymbol{\kappa}$ and allows for the diagonals to be weakly informative \citep{gelman2013bayesian}. For the spatial tuning parameter, the prior is dependent on the type of spatial data, areal or point-referenced. For areal data, the prior for $\rho$ is typically fixed at 0.99 to promote spatial smoothing, or given a uniform prior between zero and one. For point-referenced data, the prior is often uniform with bounds informed based on the expected range of the spatial variability, as in \cite{berchuck2016spatially}.

The hyperparameters for the temporal process are assigned priors in the same vein as the spatial parameters. For the $k \times k$ covariance of the latent factors, $\boldsymbol{\Upsilon} \sim \text{IW}\left(\zeta, \boldsymbol{\Omega}\right)$, with $\zeta = k + 1$ and $\boldsymbol{\Omega} = \mathbf{I}_k$. The prior for the temporal tuning parameter $\psi$ depends on the temporal correlation structure. For an AR(1) process the temporal tuning parameter $\psi$ has a transformed Beta distribution, $\psi \propto (1 + \psi)^{\gamma - 1}(1 - \psi)^{\beta - 1}$. While in the case of an exponential process, a uniform prior is more appropriate. Finally, in the case of a Gaussian likelihood, a weakly informative prior is used for the variances, $\sigma^2(\mathbf{s}_{i,o}) \sim \text{IG}(a, b)$.


\subsection{Clustering Temporal Trends}
\label{sec:clustering}

We close our introduction of the spatial factor analysis model, by describing a method for clustering regions across a spatial unit dependent on temporal change. In BNP, clustering is determined based on the posterior probability that two locations belong to the same underlying cluster, $g_j(\mathbf{s}_{i,o}, \mathbf{s}_{i',o'}) = P(\xi_{j}(\mathbf{s}_{i,o}) = \xi_j(\mathbf{s}_{i',o'}))$, which alleviates the label switching issue. Unfortunately, due to the full specification of the factor loadings matrix, none of the columns are themselves identifiable, and we can not individually cluster on the $k$ columns. Instead, we focus our clustering efforts on the factor loadings from all of the columns.

We begin by defining the loading probabilities for factor $j$ at a particular location: $\mathbf{w}_{j}(\mathbf{s}_{i,o}) = \{w_{11}(\mathbf{s}_{i,o}), \ldots, w_{1L}(\mathbf{s}_{i,o})\}$. The full set across all the latent factors is given by, $\mathbf{w}(\mathbf{s}_{i,o}) = \{\mathbf{w}_{1}(\mathbf{s}_{i,o}),\ldots, \mathbf{w}_{k^*}(\mathbf{s}_{i,o})\}$. Note that we limit to the first $k^*$ factors, which is designed to only include factors that exhibit variability across locations. We find that a good value of $k^*$ can be chosen such that, for all $j = 1,\ldots,k^*$, $\stackrel[\{i,o,i',o'\}]{}{\min}\{g_j(\mathbf{s}_{i,o}, \mathbf{s}_{i',o'})\} < 0.2$ and $\stackrel[\{i,o,i',o'\}]{}{\max}\{g_j(\mathbf{s}_{i,o}, \mathbf{s}_{i',o'})\} > 0.8$. This criteria excludes any factors that are non-informative for clustering. The final object, $\mathbf{w} = \{\mathbf{w}(\mathbf{s}_{1,1})^\top, \ldots, \mathbf{w}(\mathbf{s}_{m, 1})^\top,\ldots,\mathbf{w}(\mathbf{s}_{1, O})^\top,\ldots, \mathbf{w}(\mathbf{s}_{m, O})^\top  \}^\top$ has dimension $mO \times Lk^*$. More specifically, however, when using slice sampling each factor is truncated to $L_j^*$, and thus in theory $\mathbf{w}$ has a much smaller number of columns.

When clustering temporal trends across the spatial surface, we will use the factor loading probability matrix, $\mathbf{w}$. In particular, we apply simple k-means to $\mathbf{w}$ and use the gap-statistic to determine the proper number clusters \citep{tibshirani2001estimating}. While this process may not seem immediately intuitive, since $\mathbf{w}$ is potentially larger than the original data, we found that clustering $\mathbf{w}$ produced improved results over the raw data $\{\mathbf{Y}_1, \ldots, \mathbf{Y}_T\}$.


\subsection{Bayesian Non-parametric Prediction}
\label{sec:pred}

Once posterior samples have been obtained, prediction is often a priority. In particular,

\noindent obtaining samples from the posterior predictive distribution (PPD) is of interest, for both new spatial and temporal instances. We begin by detailing how future instances of the spatial surface can be obtained, by defining the PPD as $f\left(\mathbf{Y}_{T+1}|\mathbf{Y}\right)$. We express the PPD as an integral $\int_{\Omega} f\left(\mathbf{Y}_{T+1}|{\Omega},\mathbf{Y}\right) f\left({\Omega}|\mathbf{Y}\right) d{\Omega}$ and then further partition the integral,
\begin{equation} \label{eq:prediction}
\int_{\Omega} \underbrace{f\left(\mathbf{Y}_{T+1}|g^{-1}\left(\boldsymbol{\vartheta}_{T+1}\right),\boldsymbol{\zeta}_{T + 1}\right)}_{1} \underbrace{f\left(\boldsymbol{\eta}_ {T+1}|\boldsymbol{\eta},\boldsymbol{\Upsilon},\psi\right)}_{2} \\
\underbrace{f\left(\boldsymbol{\Lambda},\boldsymbol{\eta},\boldsymbol{\zeta},\boldsymbol{\Upsilon},\psi|\mathbf{Y}\right)}_{3} d{\Omega},
\end{equation}
where ${\Omega}=(\boldsymbol{\vartheta}_{T+1},\boldsymbol{\zeta}_{T + 1},\boldsymbol{\Lambda}, \boldsymbol{\eta},\boldsymbol{\zeta},\boldsymbol{\Upsilon},\psi)$.
The convenient form of Equation \ref{eq:prediction} is a function of three known densities that are defined as a consequence of the methodology introduced in Section \ref{sec:psbp}. As such, the PPD can be obtained by composition sampling.

Equation \ref{eq:prediction}.1 represents the observed likelihood function written in vector form and is problem specific (or in scalar form: $\prod_{o = 1}^O\prod_{i=1}^mf(Y_{T+1}(\mathbf{s}_{i,o})|g^{-1}\left(\vartheta_{T+1}(\mathbf{s}_{i,o})\right),\zeta_{T+1}(\mathbf{s}_{i,o}))$. Equation \ref{eq:prediction}.2 depends on properties of the conditional multivariate Gaussian density, yielding $f\left(\boldsymbol{\eta}_{T+1} |\boldsymbol{\eta}, \boldsymbol{\Upsilon}, \psi\right) \sim \text{MVN}\left( \mathbb{E}_{\boldsymbol{\eta}_{T+1}}, \mathbb{C}_{\boldsymbol{\eta}_{T+1}} \right).$ The moments are $\mathbb{E}_{\boldsymbol{\eta}_{T+1}} = (\mathbf{H}^+ \otimes \mathbf{I})\boldsymbol{\eta}$ and $\mathbb{C}_{\boldsymbol{\eta}_{T+1}} = \mathbf{H}^* \otimes \boldsymbol{\Upsilon}$ with $\mathbf{H}^+=[\mathbf{H}(\psi)]_{T+1,1:T} [\mathbf{H}(\psi)]_{1:T,1:T}^{-1}$ and finally $\mathbf{H}^*=[\mathbf{H}(\psi)]_{T+1,T+1} - \mathbf{H}^+[\mathbf{H}(\psi)]_{1:T,T+1}$. Here $\mathbf{H}(\psi)$ represents the temporal correlation matrix including the new time point $T + 1$, so that $[\mathbf{H}(\psi)]_{T+1,1:T}$ is a subset including the row $T+1$ and columns 1 up to $T$. Finally, Equation \ref{eq:prediction}.3 is the posterior distribution obtained in the MCMC sampler from the original model fit. Full details of the prediction theory and an extension to predicting at new spatial locations is given the Supplementary Materials online.


\section{SIMULATION EXPERIMENTS}
\label{sec:sims}


\subsection{Justifying the Spatial PSBP for Factor Analysis}
\label{sec:sim1}

In our first simulation experiment, we aimed to illuminate the importance of the spatial 

\noindent PSBP and multiplicative gamma process shrinkage priors in the presence of spatial variability. We simulated data using the full model across various settings, including a different number of latent factors, $k = 1,3,6$, and the presence or absence of spatial correlation. To simulate data from a spatial process the spatial covariance, $\mathbf{F}(\rho)$, was set to a proper CAR prior, with $\rho = 0.99$. To simulate data with no spatial dependence, the spatial covariance was fixed at the identity matrix, (i.e., $\mathbf{F}(\rho) = \mathbf{I}$). This yielded six simulation settings.

The simulation aimed to imitate the monitoring of glaucoma progression, so we fixed the number of spatial locations, ($m = 52, O = 1$), to the number on a visual field (Details of this setting follow in Section \ref{sec:glaucoma}). The visits are uniformly spaced between zero and one with total number of visits set to the average in our visual field data ($T = 10$). Furthermore, we used the adjacency matrix from the visual field, where two locations $i$ and $i'$ are considered neighbors if they share an edge or corner, $w_{ii'} = 1(i \sim i')$. Finally, we set $\boldsymbol{\kappa} = 1$, $\tau_j = 1$, $\psi = 0.3$, and $\boldsymbol{\Upsilon}$ was sampled from its prior distribution and was dependent on $k$. For each setting, 100 datasets were simulated, where every dataset is generated from one simulated instance of $\boldsymbol{\alpha}$ to ensure that the results are not affected by a particular realization. 

We now describe specific details of the model implementation, which, unless otherwise noted, apply to all subsequent modeling examples. For the spatial process, we used a proper CAR with $\rho = 0.99$ to encourage spatial dependency, similar to how the data was simulated. An exponential correlation structure was used for the temporal process, so that $\psi \sim \text{Uniform}(a_\psi, b_\psi)$. The bounds for $\psi$ cannot be specified arbitrarily since it is important to account for temporal range. We specified the following conditions for finding the bounds, $[a_{\psi}:[\mathbf{H}(a_{\psi})]_{t,t'}=0.95 , |x_t-x_{t'}| = x_{\text{max}}]$ and $[b_{\psi}:[\mathbf{H}(b_{\psi})]_{t,t'}=0.01 , |x_t-x_{t'}| = x_{\text{min}}]$, where $x_{\text{min}}$ and $x_{\text{max}}$ are the minimum and maximum temporal differences between visits. The remaining priors come directly from Section \ref{sec:hypers}, however, because there was only one spatial observation type, we specified the following prior, $\boldsymbol{\kappa} \sim \text{IG}(0.001, 0.001)$. Finally, to promote shrinkage from the gamma process prior on the columns of the factor loadings matrix, we specified, $a_1 = 1$ and $a_2 = 20$. Inference proceeds using the MCMC sampler described in Section \ref{sec:comp}, with non-convergence evaluated primarily through examination of traceplots, but also the Geweke statistic \citep{geweke1992}. We call this method Model 1.

In order to compare our introduced methodology, we compared it to various simplifications. Model 2 removed the spatial component, setting $\mathbf{F}(\rho) = \mathbf{I}$. Model 3 removed the gamma shrinkage prior, instead using independent priors, $\delta_h \sim \text{Ga}(a_1, a_2)$. Models 4 and 5 replaced the PSBP prior with a standard multivariate CAR prior for each column of the factor loadings matrix, comparable to the model of \cite{lopes2008spatial}. Furthermore, Models 4 and 5 removed the multiplicative gamma process shrinkage prior, instead using the same criteria as Model 3. Finally, Model 5 removed spatial dependency, using the identity matrix. All models were fit assuming six underlying latent factors (i.e., $k = 6$).

\setlength{\tabcolsep}{5pt}
\begin{table}[t]
\caption{Assessing the performance of the spatial stick-breaking process and multiplicative gamma process shrinkage prior. Simulation settings are defined by the number of true underlying factors ($k = 1, 3, 6$) and whether spatial dependency is present (Y: $\mathbf{F}(0.99)$, N: $\mathbf{I}$). Each of the simulated datasets has 10 uniform time points, which is the average in our visual field dataset (i.e., $T = 10$). Model fit is assessed using widely applicable information criterion (WAIC) and prediction performance is defined as accuracy of predicting the 13$^{th}$ time point, and is determined by the continuous ranked probability score (CRPS). Smaller values are preferred for both. Each summary is based on 100 simulated datasets. \label{tab:sim1}}
\centering
\begin{tabular}{ccrrrrrrrrrrrr}
  \hline
\multicolumn{2}{c}{} & & \multicolumn{5}{c}{WAIC} & & \multicolumn{5}{c}{CRPS} \\ \cline{4-14}
Space & $k$ & & M1 & M2 & M3 & M4 & M5 & & M1 & M2 & M3 & M4 & M5 \\ \hline
Y & 1 & & -1215 & -997 & -1212 & -1203 & -1058 & & 0.408 & 0.465 & 0.409 & 0.413 & 0.416 \\ 
& 3 & & -1185 & -972 & -1182 & -1006 & -875 & & 1.126 & 1.152 & 1.136 & 1.168 & 1.180 \\  
& 6 & & -1073 & -890& -1076 & -922 & -804 & & 1.622 & 1.645 & 1.610 & 1.639 & 1.648 \\ \hline
N & 1 & & -1219 & -1013 & -1219 & -1195 & -1073 & & 0.409 & 0.463 & 0.409 & 0.412 & 0.412 \\ 
& 3 & & -1178 & -966& -1178 & -1044 & -884 & & 1.087 & 1.112 & 1.094 & 1.126 & 1.134 \\ 
& 6 & & -1050 & -842 & -1061 & -907 & -789 & & 1.645 & 1.675 & 1.671 & 1.682 & 1.686 \\ \hline
\end{tabular}
\end{table}

We compared the five models using both a model fit and prediction summary. Model fit was assessed using widely applicable information criterion (WAIC) and prediction performance was defined as accuracy of predicting a 13$^{th}$ simulated time point, and was measured by the continuous ranked probability score (CRPS) \citep{hersbach2000decomposition,vehtari2017practical}. Smaller values are preferred for both.

Results are found in Table \ref{tab:sim1}. We begin by studying the results for model fit. The most clear conclusion is that Models 1 and 3, the models that have the spatial PSBP (Model 3 loses the multiplicative gamma process shrinkage prior), perform the best across all settings. The difference between Models 1 and 3 is minimal, but inclusion of the gamma shrinkage prior (Model 1) does normally fit better. The only time Model 3 outperforms Model 1 is when the true number of latent factors is the same as the simulated data (i.e., $k = 6$), indicating that the gamma shrinkage prior is useful when the true number of factors is less than the number specified in the model. Another valuable comparison is with Models 4 and 5, which do not use the stick-breaking construction or the gamma shrinkage prior (Model 5 also does not include space), and have worse performance. Clearly, in the presence of space the spatial PSBP is crucial for model fit. The prediction results mirror the same trend as the model fit, with Models 1 and 3 being superior.


\subsection{Clustering using the Spatial PSBP}
\label{sec:sim2}

In our second simulation study, we aimed to demonstrate the use of the spatial PSBP to cluster temporal changes across space. In order to do this we used a similar data generating process as the simulation in Section \ref{sec:sim1}, however we made some key changes. In particular, we simulated data based on two true clusters, where the first cluster represented the region of the visual field called the inferior nasal, which includes eight spatial locations. The second cluster consisted of the remaining locations on the visual field.

Data was generated from point-wise logistic regression models where the intercepts and slopes were drawn jointly from a spatial process, $\text{N}_{2m}\left((\boldsymbol{\beta}_0^\top, \boldsymbol{\beta}_1^\top)^\top, \boldsymbol{\kappa} \otimes \mathbf{F}(\rho)\right)$. Here, $\boldsymbol{\kappa}$ was a $2 \times 2$ dimensional covariance, with entries, $[\boldsymbol{\kappa}]_{11} = 4, [\boldsymbol{\kappa}]_{21} = -0.5$, and $[\boldsymbol{\kappa}]_{22} = 2$. Both the intercept ($\boldsymbol{\beta}_0$) and slope ($\boldsymbol{\beta}_1$) were piece-wise constant, with the components corresponding to the second cluster equal to $\beta_0$ and $\beta_1$, respectively, and the first cluster, $\beta_0 + \delta_{\beta_0}$ and $\beta_1 + \delta_{\beta_1}$. Once the intercepts and slopes had been generated across the visual field they were used in point-wise regressions, where the mean squared error (MSE) was set to $\sigma^2$ in the second cluster, and $\sigma^{2*} = \sigma^2 + \delta_{\sigma^2}$ in the first cluster.  Our simulation settings looked at varying magnitudes of $\delta_{\beta_0}$, $\delta_{\beta_1}$ and $\delta_{\sigma^2}$, which dictated variability across clusters. We set $\beta_0 = -8$, $\beta_1 = -4$, and $\sigma^2 = 3$, and then looked at $\delta_{\beta_0} = 0,6$, $\delta_{\beta_1} = 0, 3, 6$ and $\delta_{\sigma^2} = 0, 2$. We allowed for both spatial (Y: $\rho = 0.99$) and independent (N: $\rho = 0$) processes. 

\setlength{\tabcolsep}{4pt}
\begin{table}[t]
\caption{Assessing the clustering performance of the spatial PSBP and multiplicative gamma process shrinkage prior. Simulations are based on a true setting with two clusters. The first cluster is generated from a linear regression model with mean values of intercept, slope, and variance, $\beta_0 = -8$, $\beta_1 = -4$, and $\sigma^2 = 3$. The second cluster is simulated from the following mean parameters, $\beta_0^* = \beta_0 + \delta_{\beta_0}$, $\beta_1^* = \beta_1 + \delta_{\beta_1}$, and $\sigma^{2*} = \sigma^2 + \delta_{\sigma^2}$, where $\delta_{\beta_0}$, $\delta_{\beta_1}$ and $\delta_{\sigma^2}$ dictate the variability across clusters. Furthermore, model parameters were simulated either from an independent ($\rho = 0$) or spatial ($\rho = 0.99$) process. The PSBP clustering from Section \ref{sec:clustering} was compared with k-means performed on the raw data. We present the ratio of between sum of squares (SS) over total SS for the PSBP ($\text{SS}_{PSBP}$) and also the SS ratio with k-means on the raw data ($\text{SS}_{Ratio} = \text{SS}_{PSBP}/\text{SS}_{Raw}$). Greater values of $\text{SS}_{PSBP}$ are preferred, while $\text{SS}_{Ratio}$ greater than one indicate improved performance of the PSBP method. \label{tab:sim2}}
\centering
\begin{tabular}{rrrcrrrcrrrcrrrcrrr}
  \hline
\multicolumn{3}{c}{} && \multicolumn{7}{c}{Spatial Dependency} && \multicolumn{7}{c}{Independent}\\
\multicolumn{3}{c}{} && \multicolumn{3}{c}{$\text{SS}_{PSBP}$} && \multicolumn{3}{c}{$\text{SS}_{Ratio}$} && \multicolumn{3}{c}{$\text{SS}_{PSBP}$} && \multicolumn{3}{c}{$\text{SS}_{Ratio}$} \\ \cline{5-11} \cline{13-19}
$\delta_{\beta_0}$ & $\delta_{\beta_1} $ & $\delta_{\sigma^2}$ && M1 & M2 & M3 && M1 & M2 & M3 && M1 & M2 & M3 && M1 & M2 & M3 \\ 
  \hline
0 & 0 & 0 && 0.02 & 0.02 & 0.02 && 1.16 & 1.33 & 1.16 && 0.03 & 0.03 & 0.03 && 1.19 & 1.19 & 1.23 \\ 
  6 & 0 & 0 && 0.82 & 0.86 & 0.81 && 1.45 & 1.52 & 1.44 && 0.75 & 0.80 & 0.73 && 1.37 & 1.47 & 1.35 \\ 
  0 & 3 & 0 && 0.10 & 0.10 & 0.09 && 1.21 & 1.25 & 1.19 && 0.10 & 0.08 & 0.10 && 1.02 & 0.86 & 0.97 \\ 
  6 & 3 & 0 && 0.81 & 0.84 & 0.82 && 1.25 & 1.30 & 1.26 && 0.78 & 0.85 & 0.78 && 1.27 & 1.37 & 1.26 \\ 
  0 & 6 & 0 && 0.41 & 0.33 & 0.40 && 1.87 & 1.50 & 1.82 && 0.37 & 0.30 & 0.36 && 1.73 & 1.33 & 1.70 \\ 
  6 & 6 & 0 && 0.82 & 0.80 & 0.81 && 1.14 & 1.12 & 1.13 && 0.79 & 0.82 & 0.78 && 1.15 & 1.19 & 1.13 \\ 
  0 & 0 & 2 && 0.02 & 0.02 & 0.02 && 0.83 & 0.90 & 0.81 && 0.03 & 0.02 & 0.03 && 0.78 & 0.74 & 0.79 \\ 
  6 & 0 & 2 && 0.81 & 0.87 & 0.81 && 1.46 & 1.56 & 1.45 && 0.73 & 0.80 & 0.73 && 1.38 & 1.51 & 1.38 \\ 
  0 & 3 & 2 && 0.07 & 0.07 & 0.08 && 0.79 & 0.80 & 0.82 && 0.07 & 0.06 & 0.07 && 0.83 & 0.76 & 0.83 \\ 
  6 & 3 & 2 && 0.81 & 0.86 & 0.82 && 1.27 & 1.36 & 1.29 && 0.77 & 0.84 & 0.77 && 1.24 & 1.35 & 1.24 \\ 
  0 & 6 & 2 && 0.32 & 0.28 & 0.32 && 1.49 & 1.28 & 1.47 && 0.26 & 0.24 & 0.26 && 1.17 & 1.06 & 1.17 \\ 
  6 & 6 & 2 && 0.82 & 0.86 & 0.82 && 1.16 & 1.23 & 1.17 && 0.78 & 0.83 & 0.78 && 1.13 & 1.21 & 1.13 \\ 
   \hline
\end{tabular}
\end{table}

We compared the clustering technique introduced in Section \ref{sec:clustering} to a simplified version that performed k-means on the raw data. For comparison, we used the ratio of between sum of squares (BSS), $\sum_{o = 1}^O\sum_{i = 1}^m (\hat{\mathbf{w}}(\mathbf{s}_{i,o}) - \bar{\mathbf{w}})^2$, over total SS (TSS), $\sum_{o=1}^O\sum_{i=1}^m (\mathbf{w}(\mathbf{s}_{i,o}) - \bar{\mathbf{w}})^2$, where $\hat{\mathbf{w}}(\mathbf{s}_{i,o})$ represented a fitted cluster mean, so if $\mathbf{w}(\mathbf{s}_{i,o})$ belonged to cluster $c$, $\hat{\mathbf{w}}(\mathbf{s}_{i,o})= \sum_{o = 1}^O\sum_{i = 1}^m\mathbf{w}(\mathbf{s}_{i,o})1\{\mathbf{w}(\mathbf{s}_{i,o}) \in c\} / \sum_{o = 1}^O\sum_{i = 1}^m1\{\mathbf{w}(\mathbf{s}_{i,o}) \in c\}$. The quantity $\bar{\mathbf{w}}$ was the overall mean, $\sum_{o,i}\mathbf{w}(\mathbf{s}_{i,o}) / (Om)$. For adequate clusters, we would have expected this ratio to be close to one, because a large BSS indicates high variability between clusters (and accordingly, small variability within clusters). We presented the SS ratio (BSS / TSS) for the PSBP clustering ($\text{SS}_{PSBP}$). In place of the SS ratio for the raw clustering ($\text{SS}_{Raw}$), we presented the ratio $\text{SS}_{Ratio} = \text{SS}_{PSBP}/\text{SS}_{Raw}$, for comparison purposes. For $\text{SS}_{Ratio}$, values greater than one indicate improved performance of the PSBP method. We presented results only for Models 1-3, since they are the only models with clustering capabilities.

The results of the simulation can be found in Table \ref{tab:sim2}. By presenting both $\text{SS}_{PSBP}$ and $\text{SS}_{Ratio}$ we can interpret the clustering performance in absolute and relative terms. In general, we can see that, in relative terms, the PSBP has improved clustering performance. In the only settings where the raw clustering technique has better relative performance ($\delta_{\beta_0} = 0$, $\delta_{\beta_1} = 0 \text{ or }3$ and $\delta_{\sigma^2} = 2$), the results are negligible, because the $\text{SS}_{PSBP}$ are close to zero, meaning the settings were overly difficult. Overall, it appears that the biggest boost in performance for the PSBP is when the true underlying intercepts are different between groups. There is also evidence that the PSBP process is capable of detecting clusters based on differences in the underlying true slope (i.e., $\delta_{\beta_1}$), which is particularly impactful for clustering spatial locations based on temporal trajectories. 


\section{DATA ILLUSTRATIONS}
\label{sec:data}


\subsection{Glaucoma Progression using Visual Fields}
\label{sec:glaucoma}

In our first case study using real data, we used the spatial PSBP to determine glaucoma progression from longitudinal visual fields. Glaucoma, an optic neuropathy, is the leading cause of irreversible vision loss worldwide. Although glaucomatous damage is irreversible, early treatment can usually prevent or slow down progression to functional damage and visual impairment. Estimation of rates of functional deterioration by visual fields is essential for determining patient prognosis and aggressiveness of therapy \citep{weinreb2014pathophysiology}.

\begin{figure}[t]
\begin{center}
\includegraphics[scale = 0.63]{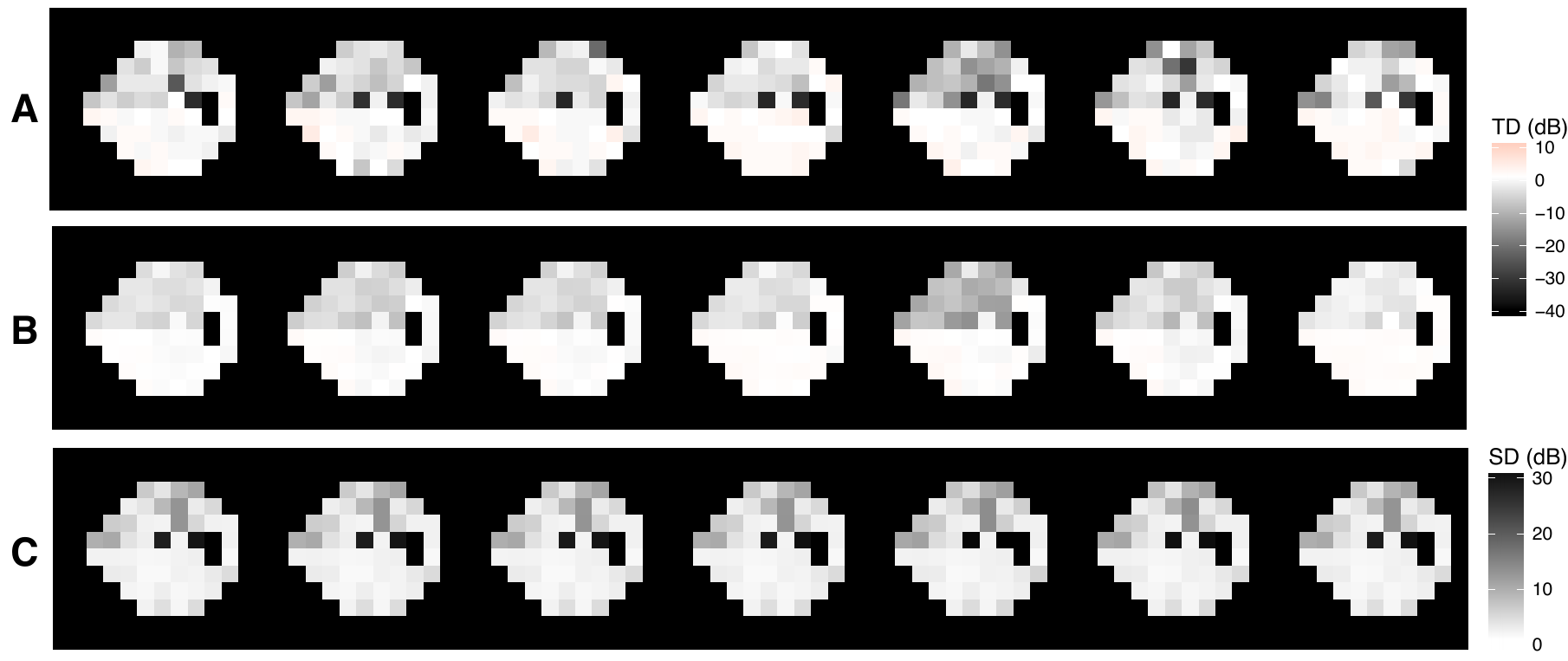}
\caption{Example longitudinal series of visual fields, presented in total deviation (TD), a measure of age-adjusted loss. Negative values of TD indicate poorer vision. Clinicians are tasked with determining wether the rate of progression is clinically significant for intervention. \label{fig:fig1}}
\end{center}
\end{figure}

Visual fields are a psychophysical procedure that assesses a patient's field of vision, with standard automated perimetry (SAP) being the default method. In this study, we analyzed fields generated from the Humphrey Field Analyzer-II (HFA-II; Carl Zeiss Meditec Inc., Dublin, CA). The HFA-II is an interactive technology that assesses a patient's reaction as light is systematically introduced at gridded locations across their visual field. In this study, we represented functional loss using total deviation (TD) values, an age-adjusted measure of sensitivity loss, measured in decibels (dB). TD is a continuous measure, with large negative values indicating functional loss. An example longitudinal series of visual fields can be found in Figure \ref{fig:fig1}A. Our data included 79 patients (110 eyes) diagnosed with glaucoma at baseline, with an average of 10 clinic visits and 4 years of follow-up. Of the 110 glaucomatous eyes, 51 (46\%) were defined as progressing and the remaining 59 (54\%) were stable. More details about the glaucoma population are given in the online Supplementary Materials. See \cite{berchuck2019diagnosing} for a more in depth introduction to visual fields. 

\subsubsection{Rates of Glaucoma Progression}
\label{sec:diag}

The spatial factor analysis was fit to each of the 110 eyes in our study. For the example

\noindent longitudinal series of visual fields, presented in Figure \ref{fig:fig1}, we present posterior mean and standard deviation fits at each clinic visit. From this visualization, we see that the method is properly spatially smoothing the observed data to better reveal patterns across time. 

We are interested in using the spatial factor analysis model to improve clinicians' ability to quantify rates of change across time. Our introduced methodology is appropriate for assessing longitudinal changes on the visual field, because instead of analyzing each location, it models temporal changes of underlying regions. This is much closer to how a clinician interprets change on the visual field, as glaucoma has characteristic patterns. To this end, we performed independent linear regressions of the posterior mean estimates of the latent factors across time. The two-sided p-values from these regressions were used as predictors of progression. In particular, we present five variations that include, all six of the factors (i), the first three (ii), and only the first (iii), second (iv), and third (v) factors. To assess the diagnostic ability of these methods to discriminate progression status, we performed logistic regressions of the resulting p-values of the latent factors across time, using the predicted probabilities of progression as a diagnostic.

We compared the probabilities from the latent factors to established methods of determining progression on SAP fields, mean deviation (MD) and pattern standard deviation (PSD). Both MD and PSD are age-adjusted measures of vision loss on a visual field, with MD representing a global loss and PSD indicating the level of localized loss. In practice MD (and PSD) is used to assess rates of progression using OLS regression across time, with a lower (and upper) p-value less than 0.05 indicating progression. Again, p-values were regressed against disease status and predicted progression probabilities were obtained.

\begin{figure}[t]
\begin{center}
\includegraphics[scale = 0.5, trim = 0cm 1cm 0cm 0cm]{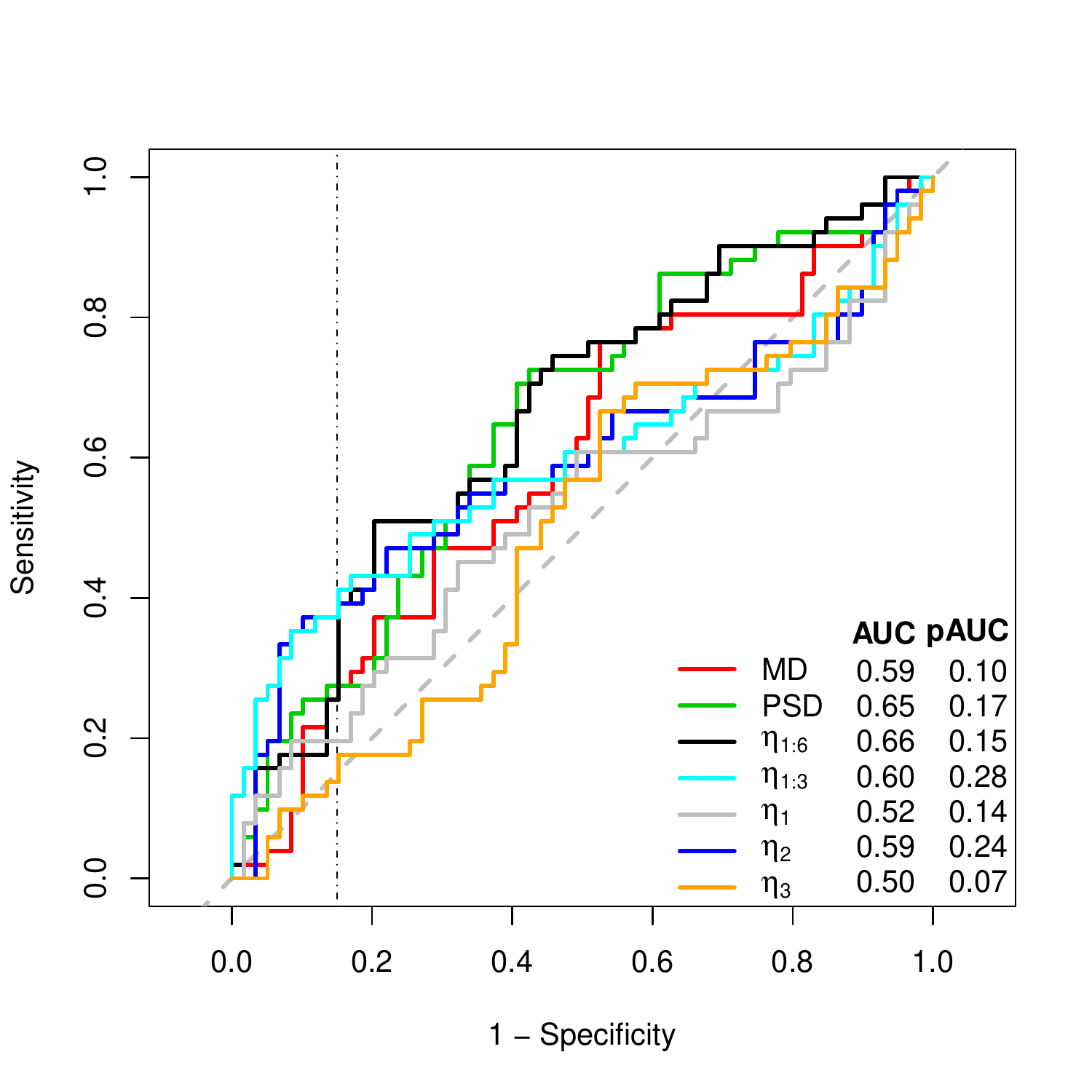}
\caption{Receiver operating characteristic (ROC) curves for metrics diagnosing structural glaucomatous progression. Established metrics mean deviation (MD) and pattern standard deviation (PSD) are compared with metrics derived from regressing the posterior latent factors across time. Also, presented are area under the ROC curve (AUC) and partial AUC (pAUC), limited to the region of specificity greater than 85\%.\label{fig:roc}}
\end{center}
\end{figure}

We compared diagnostics using area under the receiver operating characteristic (ROC) curve (AUC) and partial AUC (pAUC). Larger values of AUC and pAUC indicate superior discriminatory ability. Based on the precedent of a previous study, we limited the pAUC to regions of clinically relevant specificity, 85-100\% \citep{berchuck2019improved}. 

Of the two established methods, PSD had better performance with an AUC of 0.65 and pAUC of 0.17 (Figure \ref{fig:roc}). When all six latent factors were included in the analysis, the AUC was improved slightly, however the pAUC decreased. This is problematic, because the pAUC is clinically more meaningful than overall AUC. Through inspection of the posterior factors, we determined that for the majority of eyes only three factors contained meaningful data (a result of the shrinkage prior on the loadings matrix). When we limited the metric to only contain the first three factors, the pAUC nearly doubled with a maximum pAUC of 0.28. When further exploring each of the first three factors independently, it became clear that the meaningful information had been encoded in the second factor, as the corresponding ROC curve has a much steeper trajectory from 100\% specificity. These results indicate that the rates of change learned from the posterior latent factors have clinical utility in determining underlying structural progression from visual fields.

\subsubsection{Clustering Temporal Trends}
\label{sec:clust}

\begin{figure}[t]
\begin{center}
\includegraphics[scale = 0.65]{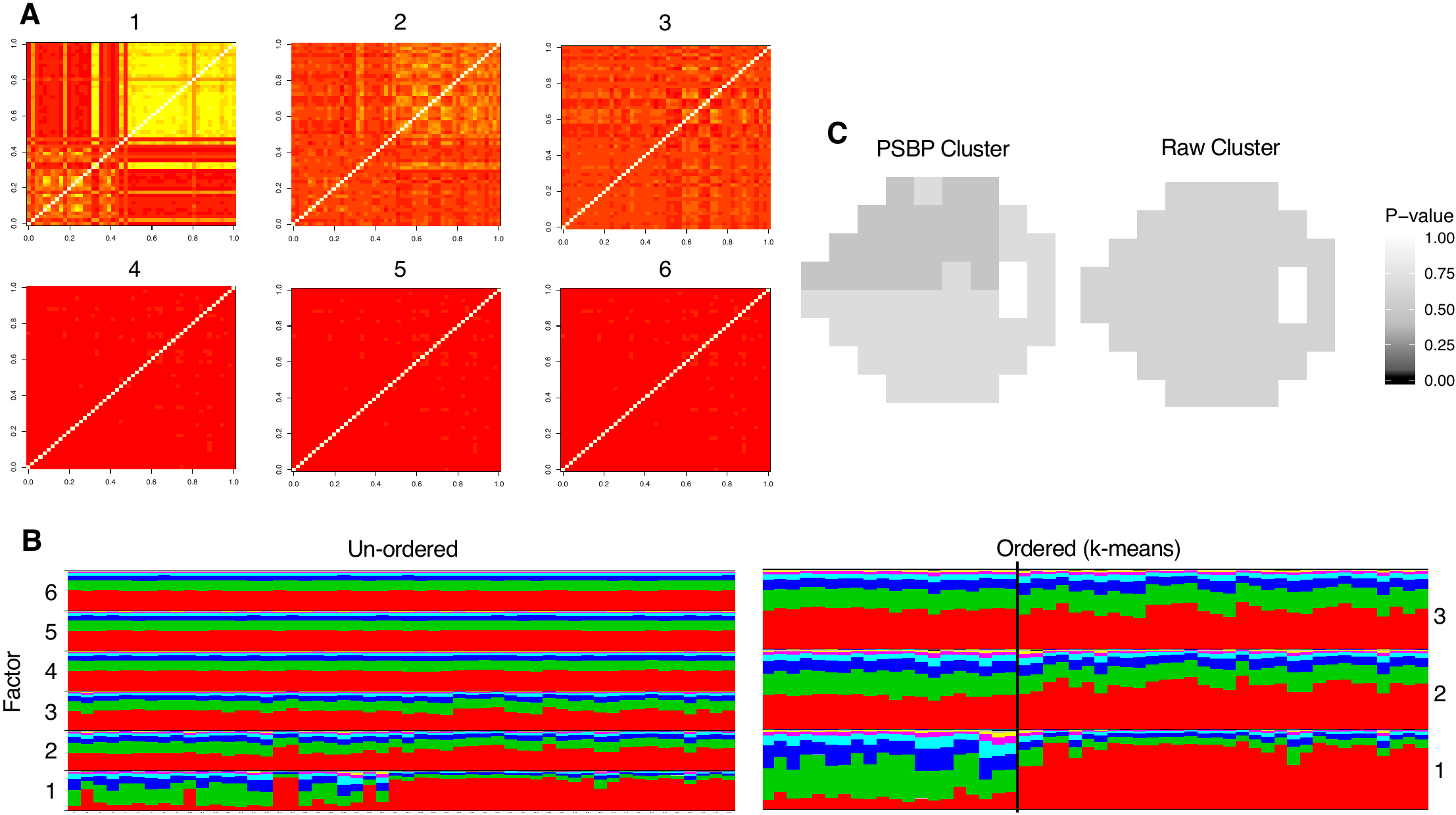}
\caption{Demonstrating the clustering ability of the spatial PSBP for the example patient in Figure \ref{fig:fig1}. Frame A contains the posterior probabilities of belonging to the same cluster, $g_j(\mathbf{s}_{i,o}, \mathbf{s}_{i',o'})$, for each factor On the left of frame B, the stacked posterior probabilities $\mathbf{w}(\mathbf{s}_{i,o})$ are presented in their original order, while on the right the ordered version is presented, with only the meaningful factors included. Finally, in frame C, clusters are presented with p-values, which are the average lower p-values from point-wise logistic regressions of the PPD across all locations. \label{fig:vfex}}
\end{center}
\end{figure}

We close this data illustration by demonstrating the clustering ability of the spatial PSBP detailed in Section \ref{sec:clustering}. In Figure \ref{fig:vfex}A, for each latent factor, the posterior probabilities of belonging to the same cluster, $g_j(\mathbf{s}_{i,o}, \mathbf{s}_{i',o'})$, are presented. For this eye, only the first three factors contain meaningful information. This is further reinforced in Figure \ref{fig:vfex}B, where the stacked posterior probabilities $\mathbf{w}(\mathbf{s}_{i,o})$ are presented across the spatial surface and observation type ($\mathbf{w}$). On the left of Figure \ref{fig:vfex}B, the probabilities are presented in their original order (i.e., un-ordered), while the right frame represents the ordered factors (according to k-means with two groups, determined using the gap statistic). The ordered version removed the three non-informative factors based on the criteria from Section \ref{sec:clustering}. Due to the identifiability issue, the right of Figure \ref{fig:vfex}B is required for clustering. 

We presented the clusters obtained from the spatial PSBP (left) and clustering of the raw data using k-means with one cluster, based on the gap statistic (right). The clusters, presented in Figure \ref{fig:vfex}C, show p-values, which are the average lower p-values from point-wise regressions of the PPD across all locations within each cluster. From this presentation, we can see that the spatial PSBP was capable of producing a map with regions of varying temporal trends that are straightforward for clinicians to interpret. While clustering the raw data was non-informative, the PSBP illustrated a region with faster temporal trajectories, that may require intervention, and corresponds with true progression (Figure \ref{fig:fig1}A).


\subsection{Malaria Incidence in Peru}
\label{sec:peru}

In our second case study using real data, we investigated malaria counts across Loreto, 

\noindent Peru's northernmost region that is located in the Amazon rainforest. This case study is a nice complement to the glaucoma example, as it introduces additional complexities, including dealing with count data, having multiple spatial observation types (i.e., malaria types), and introducing covariates into the mean process.

Count data fit nicely into the general framework introduced in Equation \ref{eq:lik}, which allows for a general likelihood specification. To model malaria counts we used conventions based on the Peruvian Ministry of Health, which defines significant levels of malaria based on observed counts from previous years. In particular, define the counts of malaria $c_t(\mathbf{s}_{i,o})$ at time $t$ for district $i$ and malaria type $o$. For malaria data, the temporal index is actually composed of two dimensions that represent year and epidemiological week, $t = \{y, w\}$, for $y = 2012, \ldots, 2018$, and $w = 1, \ldots, 52$. Furthermore, the Loreto region has 51 districts and there are two types of malaria monitored, \emph{P. falciparum} ($o = 1$) and \emph{P. vivax} ($o = 2$).

In order to model malaria counts, we defined the proportion of malaria $p_{yw}(\mathbf{s}_{i,o}) = c_{yw}(\mathbf{s}_{i,o}) / n_{yw}(\mathbf{s}_{i,o})$, where $n_{yw}(\mathbf{s}_{i,o})$ represents the population. Then, we defined our outcome as $Y_{yw}(\mathbf{s}_{i,o}) = 1\{p_{yw}(\mathbf{s}_{i,o}) > \mathbf{p}_{yw}(\mathbf{s}_{i,o})\}$, where $\mathbf{p}_{yw}(\mathbf{s}_{i,o}) = \sum_{x = (y - 5)}^{y - 1} p_{xw}(\mathbf{s}_{i,o}) / 5$, is the average proportion of cases for a malaria type at a location over the past five years. This binary definition, allowed us to model the probability of exceeding the number of cases seen in the past five years, $\pi_{yw}(\mathbf{s}_{i,o})$, an important indicator for specifying interventions.

To best model the mean process, we incorporated the following covariates through $\mathbf{x}_{t}(\mathbf{s}_{i,o})$, rainfall (millimeters), temperature (Celcius), and an indicator of being in the rainy season, which is approximately from February-July, but defined using weeks, $1(w \in \{6,\ldots,31\})$, in addition to an intercept, so that $p = 4$ \citep{vittor2009linking}. To demonstrate the spatial PSBP methodology, we used data from the entire 2017 year and the first five weeks of 2018, in order to predict malaria severity into the rainy season of 2018. This yielded the following set of temporal observations, $\{2017, 1\}, \ldots, \{2017, 52\}, \{2018, 1\}, \ldots, \{2018, 5\}$, resulting in $t = 1,\ldots,T$, with $T = 57$.

To model the malaria indicator we used a binomial likelihood, specified as follows, $Y_{t}(\mathbf{s}_{i,o}) \sim \text{Binomial}(n_{t}(\mathbf{s}_{i,o}), \pi_{t}(\mathbf{s}_{i,o}))$, where $\pi_{t}(\mathbf{s}_{i,o}) = g^{-1}(\vartheta_{t}(\mathbf{s}_{i,o}))$. Here $n_t(\mathbf{s}_{i,o})$ represents the population of each district in Loreto, which in our data analysis is fixed at one, $n_t(\mathbf{s}_{i,o}) = 1$. The likelihood is then, $\prod_{t = 1}^{T} \prod_{o = 1}^O \prod_{i = 1}^m \exp\{\vartheta_{t}(\mathbf{s}_{i,o})Y_{t}(\mathbf{s}_{i,o})\}\left(1 + \exp\{\vartheta_{t}(\mathbf{s}_{i,o})\}\right)^{-n_{t}(\mathbf{s}_{i,o})}$. While inference can proceed using this likelihood, it is computationally intensive due to a loss of conjugacy across the majority of parameters. Computation can be made feasible through data augmentation using P{\'o}lya--Gamma (PG) latent variables \citep{polson2013bayesian}.

In particular, based on \cite{polson2013bayesian}, we chose to model the observed data through a joint likelihood, $f\left(Y_{t}(\mathbf{s}_{i,o}), \omega_{t}(\mathbf{s}_{i,o}) | \vartheta_{o}(\mathbf{s}_{i,o}), n_{t}(\mathbf{s}_{i,o})\right)$, that consists of an augmented parameter with distribution, $\omega_{t}(\mathbf{s}_{i,o}) \sim\text{PG}(n_{t}(\mathbf{s}_{i,o}), 0)$. The reason for this is that the conditional distribution $f(\omega_{t}(\mathbf{s}_{i,o}) | \vartheta_{t}(\mathbf{s}_{i,o}), n_{t}(\mathbf{s}_{i,o}))$ is a tilted version of the PG with the following density, $\left[\left(1 + \exp\{\vartheta_{t}(\mathbf{s}_{i,o})\}\right) / \left(2\exp\{\vartheta_{t}(\mathbf{s}_{i,o}) / 2\}\right)\right]^{n_{t}(\mathbf{s}_{i,o})} \exp\left\{-0.5\vartheta^2_{t}(\mathbf{s}_{i,o})\omega_{t}(\mathbf{s}_{i,o})\right\} f(\omega_{t}(\mathbf{s}_{i,o}))$ using the fact that $\cosh\{x\} = (1 + \exp\{2x\}) / (2\exp\{x\})$. 

This is useful, because the likelihood can be expressed in the form of a Gaussian kernel, $\prod_{t = 1}^{T} \prod_{o = 1}^O \prod_{i = 1}^m \exp\left\{-0.5\omega_{t}(\mathbf{s}_{i,o}) \left(Y^*_{t}(\mathbf{s}_{i,o}) - \vartheta_{t}(\mathbf{s}_{i,o}) \right)^2  \right\}$, where $\chi_{t}(\mathbf{s}_{i,o}) = Y_{t}(\mathbf{s}_{i,o}) - n_{t}(\mathbf{s}_{i,o}) / 2$ and $Y^*_{t}(\mathbf{s}_{i,o}) = \chi_{t}(\mathbf{s}_{i,o}) / \omega_{t}(\mathbf{s}_{i,o})$. This can be further expressed in vector form, $\prod_{t=1}^{T}\exp\left\{- \frac{1}{2} \left(\mathbf{Y}_t^* - \mathbf{X}_t \boldsymbol{\beta} - \boldsymbol{\Lambda} \boldsymbol{\eta}_t \right)^T  \boldsymbol{\Delta}_t  \left(\mathbf{Y}_t^* - \mathbf{X}_t \boldsymbol{\beta} - \boldsymbol{\Lambda} \boldsymbol{\eta}_t \right) \right\}$, with $\boldsymbol{\Delta}_t = \text{diag}(\boldsymbol{\omega}_t)$ and $\mathbf{Y}_t^*$, $\boldsymbol{\vartheta}_t$ and $\boldsymbol{\omega}_t$ defined as vectors stacked the same as the original $\mathbf{Y}_t$. Because this is the kernel of a Gaussian, we maintain conjugacy and will only need to add an additional sampling step for $\omega_{t}(\mathbf{s}_{i,o})$. More details of this derivation are given in the Supplementary Materials.

\begin{figure}[t]
\begin{center}
\includegraphics[scale = 0.76]{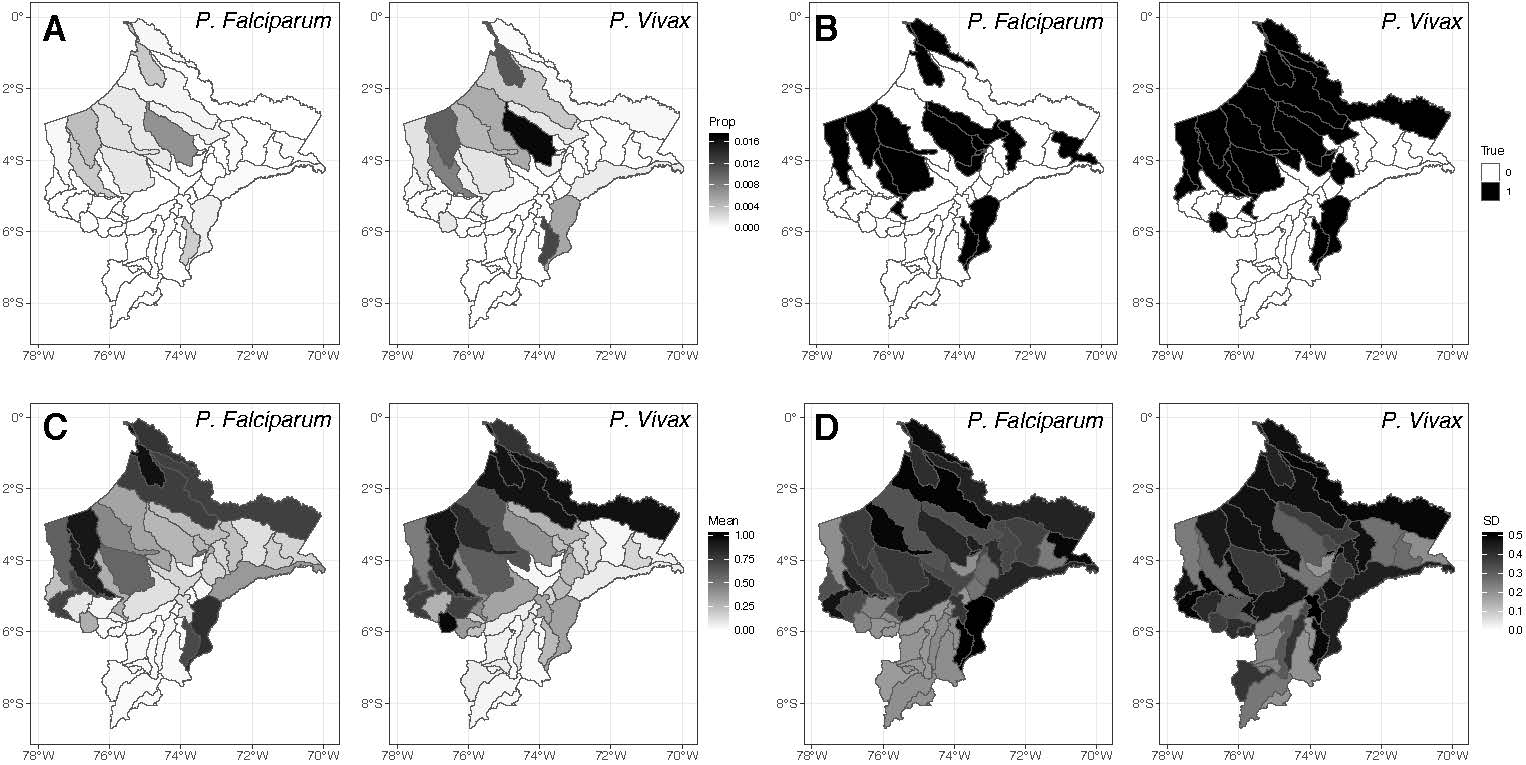}
\caption{Presenting malaria predictions for the fifth week of the 2018 rainy season. A and B show the true proportion, $p_{t}(\mathbf{s}_{i,o})$, and indicator outcomes, $Y_{t}(\mathbf{s}_{i,o})$, across Loreto for both types of malaria. In frames C and D, we present the posterior mean predicted probabilities of exceeding the average amount of malaria and their standard deviations (SD), respectively. \label{fig:mal1}}
\end{center}
\end{figure}

To demonstrate the spatial PSBP process using the malaria data, we present predictions for the rainy season in 2018 using the prediction theory from Section \ref{sec:pred}. Results are presented from predictions of the fifth week of the rainy season (i.e., t = 62 $\Longleftrightarrow$ $\{y = 2018, w = 10\}$. Figures \ref{fig:mal1}A and B show the proportion, $p_{t}(\mathbf{s}_{i,o})$, and indicator outcomes, $Y_{t}(\mathbf{s}_{i,o})$, across Loreto for both types of malaria. Then, in frames C and D of Figure \ref{fig:mal1}, we presented the posterior mean predicted probabilities of exceeding the average amount of malaria and their standard deviations (SD), respectively. While these posterior predictions are useful, they can be difficult for government agencies to draw actionable decisions. 

\begin{figure}[t]
\begin{center}
\includegraphics[scale = 0.76]{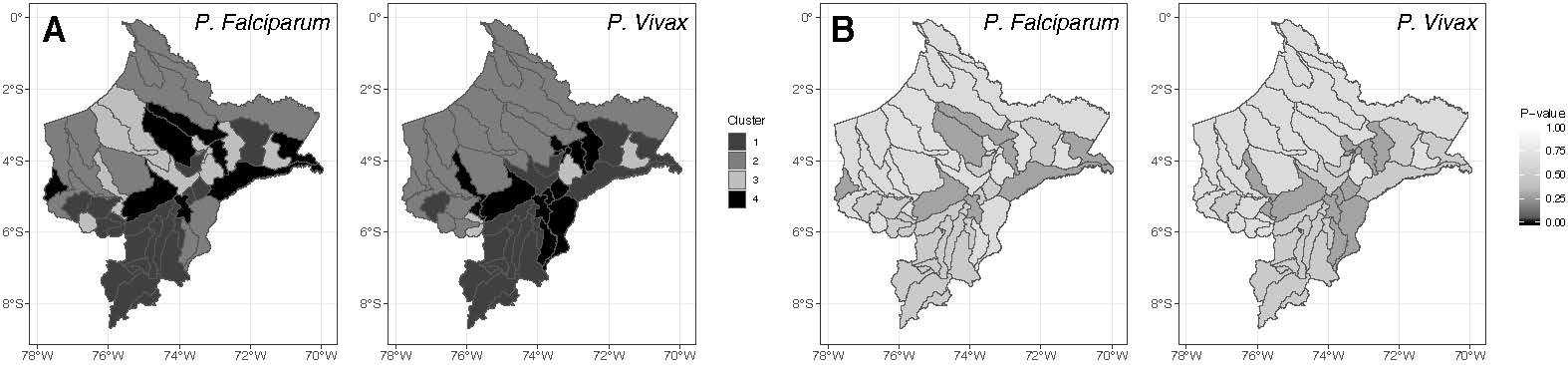}
\caption{Presenting results from applying the PSBP clustering to the rates of malaria across the Loreto region. In frame A, four clusters are presented that have been identified to have similar temporal rates of change. Frame B presents the same clusters with p-values, which are the average upper p-values from point-wise logistic regressions of the PPD across all locations and malaria types within each cluster.\label{fig:mal2}}
\end{center}
\end{figure}

Fortunately, the clustering method of the spatial PSBP, introduced in Section \ref{sec:clustering}, can be used to provide more actionable information about where to focus intervention efforts. In Figure \ref{fig:mal2}A, the resulting clusters are presented, where it can be seen that four groups were found. Recall, that clustering is based on temporal changes, so that within groups, locations across both malaria types are presumed to have similar rates of change. The clusters themselves are only informative about groupings and do not differentiate on speeds of change, so, similar to the glaucoma visual fields in Figure \ref{fig:vfex}, we presented p-values over the the clusters in Figure \ref{fig:mal2}B. The p-values are the average upper p-values from point-wise logistic regressions of the PPD across all locations and malaria types within each cluster.

The representation of malaria incidence in Figure \ref{fig:mal2}B is useful, because it allows for the Ministry of Health to focus interventions on areas across Loreto that have similarly increasing rates of malaria. For example, in Figure \ref{fig:mal1}C, we can see that for \emph{P. vivax} the most northern district has a posterior mean probability of close to one. While interventions should likely be performed in the district, based on Figure \ref{fig:mal2}B, it appears that districts centered near ($5^\circ$S, $74^\circ$W) all belong to a cluster with a p-value around 0.2 (the smallest of all the clusters). This information, not available from standard predictions, can alert the Ministry of Health to allocate resources to these districts.


\section{SUMMARY}
\label{sec:sum}

We have provided a factor analysis that can be used in the setting of spatial correlation. The spatial dependencies are introduced through a spatial BNP prior on the columns of the factor loadings matrix. The prior incorporates a multiplicative gamma process shrinkage prior to adaptively learn the proper number of latent factors. We have described an efficient MCMC sampler to accompany the model, which has been published as an \texttt{R} package, \texttt{spBFA}. We have illustrate the model's performance in both simulated and real data examples. In particular, we showed that by encoding spatial information into the loadings matrix, meaningful factors were learned that respect the observed neighborhood dependencies, making them useful for assessing rates of change across space. 

While much of the factor analysis literature performs dimension reduction on the mean process for Gaussian data, we place no such restriction, allowing for a general likelihood and non-linear relationship with the latent factors. We illustrated this through our modeling of malaria counts with a binomial likelihood. Although our specification of both space and time are separable and stationary Gaussian processes, we have noted that the resulting model is neither separable, stationary nor Gaussian. 

Using our spatial PSBP prior we were able to find regions within the spatial domain with similar temporal trajectories, an important task in many applied settings. Through simulation, we showed that our clustering technique is preferred over a standard clustering routine. In real data, we showed that aggregating rates of change at the cluster level produced patterns that aided in making actionable decisions. The task of clustering trajectories has been addressed sparsely in the literature, with the majority of methods clustering trends with no shape constraint, like our temporal process. A recent method by \cite{napier2018bayesian}, instead, clusters trajectories on pre-specified parametric forms (e.g., linear). While this technique can be limiting, in applied settings with known trajectories, it has utility and therefore would be a valuable extension to our model.


\section*{SUPPLEMENTARY MATERIAL}
\label{sec:sm}

The supplementary materials contains details related to the model, including full conditionals, moment derivations, prediction theory, and also specifics of the glaucoma population.

\bibliographystyle{References/Chicago.bst}

\bibliography{References/references.bib}

\end{document}